\documentclass{aastex}          
\usepackage{spr-astr-addons}    

\usepackage{graphicx}        
\usepackage{color}           
\usepackage{url}             
\usepackage{lscape}
\usepackage{longtable}
\usepackage{multirow}
\usepackage{rotating}

\usepackage{amsmath}

\newcommand{\etal}{{\it et al.}}

\begin{document}
%
\title{Error Analysis regarding the calculation of
Nonlinear Force-free Field}

\shorttitle{<Error Analysis regarding the calculation of NLFF
Field>} \shortauthors{<Liu, S et al.>}

\author{S. Liu\altaffilmark{1}}
\author{H.Q. Zhang\altaffilmark{1}}
\author{J.T. Su\altaffilmark{1}}

\altaffiltext{1}{Laboratory of Solar Activity, National Astronomical
Observatory, Chinese Academy of Sciences,
        Beijing, China}

\begin{abstract}
Magnetic Field extrapolation is an alternative method to study
chromospheric and coronal magnetic fields. In this paper, two
semi-analytical solutions of force-free fields (\citeauthor{low90},
\citeyear{low90}) have been used to study the errors of nonlinear
force-free (NLFF) fields based on force-free factor $\alpha$. Three
NLFF fields are extrapolated by approximate vertical integration
(AVI) \citet{son06}, boundary integral equation (BIE) \citet{yan00}
and optimization (Opt.) \citet{wie04} methods. Compared with the
first semi-analytical field, it is found that the mean values of
absolute relative standard deviations (RSD) of $\alpha$ along field
lines are about 0.96-1.19, 0.63-1.07 and 0.43-0.72 for AVI, BIE and
Opt. fields, respectively. While for the second semi-analytical
field, they are about 0.80-1.02, 0.67-1.34 and 0.33-0.55 for AVI,
BIE and Opt. fields, respectively. As for the analytical field, the
calculation error of $\langle|RSD|\rangle$ is about 0.1 $\sim$ 0.2.
It is also found that RSD does not apparently depend on the length
of field line. These provide the basic estimation on the deviation
of extrapolated field obtained by proposed methods from the real
force-free field.
\end{abstract}

\section{Introduction}
The magnetic field plays key roles in a variety of dynamical
processes, particularly in eruptive phenomena such as filament
eruptions, flares and coronal mass ejections. The topological
structure is an important properties of spatial magnetic field.
Therefore it is important to understand three-dimensional properties
of magnetic field in order to elaborate efficient theoretical
models, which will contribute to understand solar activity. At
present, due to the restrictions of observational technique, we can
not get the accurate information on magnetic field in the hot and
tenuous corona. This can only be done at the level of the cooler and
denser photosphere. In spite of that, the extrapolation of
magnetograms, measured maps of the photospheric or low chromospheric
field, into the corona is the most important technique to supply the
missing information about the spatial magnetic field. The field
extrapolation is based on force-free assumption (\citeauthor{aly89},
\citeyear{aly89}), which assumes that the corona is free of Lorentz
forces. For the low-$\beta$ corona where the plasma is tenuous, the
force-free assumption is appropriate. The force-free field obeys the
following equations:
\begin{equation}
\label{wlfc}
 \nabla \times \textbf{B} = \alpha(\textbf{r})
\textbf{B},
\end{equation}
\begin{equation}
\label{wstj}
 \nabla \cdot \textbf{B} = 0,
\end{equation}
where $\alpha$ is called force-free factor. If $\alpha$ = 0, the
equations represent a potential field (a current-free field). If
$\alpha$ = constant, they describe a current-carrying linear
force-free (LFF) field, and if $\alpha$ = $f(\textbf{r})$ it is a
general NLFF field. This two equations also imply that:
\begin{equation}
\label{nabla_alpha} \textbf{B}\cdot\nabla \alpha  = 0.
\end{equation}
Equation ($\ref{nabla_alpha}$) demonstrates that $\alpha$ is
invariant along field lines of $\textbf{B}$. The scalar $\alpha$ is
in general a function of space and identifies how much current flows
along each field line. Thus, the deviations of $\alpha$ along one
field line at some extent can demonstrate performance of field
extrapolation.

At present, the potential ($\alpha$ = 0) and linear force-free
($\alpha$ = constant) field extrapolations reached a mature
development, but they describe the magnetic field above the
photosphere in a very restricted manner. According to the practice
conditions of solar magnetic field, it is not a current-free field
and NLFF should be more reasonable. Recently, several NLFF field
extrapolation methods have been proposed ($e.g.$, \citeauthor{wu90},
\citeyear{wu90}; \citeauthor{cuper90}, \citeyear{cuper90};
\citeauthor{demo92}, \citeyear{demo92}; \citeauthor{mic94},
\citeyear{mic94}; \citeauthor{rou96}, \citeyear{rou96};
\citeauthor{ama97}, \citeyear{ama97}; \citeauthor{sak81},
\citeyear{sak81}; \citeauthor{cho81}, \citeyear{cho81};
\citeauthor{yan00}, \citeyear{yan00}; \citeauthor{whe00},
\citeyear{whe00}; \citeauthor{wie04}, \citeyear{wie04};
\citeauthor{son06}, \citeyear{son06}; \citeauthor{he08},
\citeyear{he08}). The performance of above extrapolation methods
have been studied in several papers ($e.g.,$ \citeauthor{sch06},
\citeyear{sch06}; \citeauthor{ama06}, \citeyear{ama06};
\citeauthor{regnier04}, \citeyear{regnier04}; \citeauthor{yan06},
\citeyear{yan06}; \citeauthor{wie06}, \citeyear{wie06}
\citeauthor{valo07}, \citeyear{valo07}; \citeauthor{regnier07},
\citeyear{regnier07}; \citeauthor{sch09}, \citeyear{sch09};
\citeauthor{he11}, \citeyear{he11} and \citeauthor{lius11}
\citeyear{lius11} ), comparing to models or observations ($e.g.$,
the appearance of extreme ultraviolet (EUV) and X-ray or loops)
those methods above mentioned can give reasonable results at the
level of macroscopic structure.

Since the field extrapolation is an important technique to study the
coronal magnetic field, the performance of field extrapolation
become special importance. From Equation ($\ref{nabla_alpha}$), it
can been seen that though $\alpha$ may change from one field line to
another, it must be a constant along one field line. Knowing
$\alpha$ is an important parameter worthy to be researched, in this
paper the distributions of $\alpha$ between extrapolated field and
semi-analytical field are compared; the calculation error of
$\alpha$ is discussed; the performance of $\alpha$ along field line
are studied specially.

The paper is organized as follows: firstly, the description of
extrapolation methods and semi-analytical field will be introduced
in Section~\ref{S-Alg}; secondly, the results of distribution
comparison and calculation error of force-free factor $\alpha$ are
shown in Section~\ref{S-Result}; at last, the short discussions and
conclusions will be given in section~\ref{S-Conl}.

\section{Extrapolation methods and Semi-analytical field}
\label{S-Alg}

\subsection{Approximate vertical integration method}

The approximate vertical integration (AVI) method
(\citeauthor{son06}, \citeyear{son06}) was improved from the
vertical integration proposed by \citeauthor{wu90}
(\citeyear{wu90}). In this method, it is assumed that the magnetic
field components is given by the following formula,
\begin{equation}
\label{q-avi1} \textbf{B}_{x} = \xi_{1}(x,y,z)F_{1}(x,y,z),
\end{equation}
\begin{equation}
\textbf{B}_{y} = \xi_{2}(x,y,z)F_{2}(x,y,z),
\end{equation}
\begin{equation}
\label{q-avi2} \textbf{B}_{z} = \xi_{3}(x,y,z)F_{3}(x,y,z),
\end{equation}
assuming the second-order continuous partial derivatives in a
certain height range, 0$ <z<$H (H is the calculated height from the
photospheric surface). In Equations (\ref{q-avi1})-(\ref{q-avi2}),
$\xi_{1}, \xi_{2}$ and $\xi_{3}$ mainly depend on $z$ and slowly
vary with $x$ and $y$, while $F_{1}, F_{2}$ and $F_{3}$ mainly
depend on $x$ and $y$ and weakly vary with $z$. After constructing
the magnetic field, the following integration equations,

\begin{equation}
\label{a}
 \dfrac{\partial B_{x}}{\partial z} =
 \dfrac{\partial B_{z}}{\partial x} + \alpha B_{y},
\end{equation}
\begin{equation}
 \dfrac{\partial B_{y}}{\partial z} =
 \dfrac{\partial B_{z}}{\partial y} - \alpha B_{x},
\end{equation}
\begin{equation}
 \dfrac{\partial B_{z}}{\partial z} =
 -\dfrac{\partial B_{x}}{\partial x} -
 \dfrac{\partial B_{y}}{\partial y},
\end{equation}
\begin{equation}
\label{b}
 \alpha B_{z} = \dfrac{\partial B_{y}}{\partial x}
  - \dfrac{\partial B_{x}}{\partial y},
\end{equation}
can be used to calculate extrapolated field. There exists
singularity problem due to differential operation for AVI method,
hence we smooth data locally where exists singularity ({\it cf.}
\citeauthor{lius11}, \citeyear{lius11}).
\subsection{Boundary integral equation method}
The boundary integral equation (BIE) method proposed by
\citeauthor{yan00} (\citeyear{yan00}), which uses the integration
function to extrapolate the magnetic field. In this method, an
optimized parameter $\lambda$, which is the function of spatial
position $\textbf{x}$, must be found through iteration. The integral
\begin{equation}
\label{dbie} \textbf{B}(x_{i}, y_{i}, z_{i}) =
\int_\Gamma\dfrac{z_{i}[\lambda r \sin(\lambda r)+\cos(\lambda r)]
\textbf{B}_{0}(x, y, 0)}
{2\pi[(x-x_{i})^2+(y-y_{i})^2+z_{i}^2]^{3/2}},
\end{equation}
is used to calculate the magnetic field, where $r$ =
$[(x-x_{i})^2+(y-y_{i})^2+z_{i}^2]^{1/2}$ and $\textbf{B}_{0}$ is
the magnetic field of photospheric surface. This method is to find
suitable values of $\lambda$ through iteration and try to make sure
that the extrapolated field is force-free and divergence-free ({\it
cf.} \citeauthor{he08}, \citeyear{he08} and \citeauthor{li04},
\citeyear{li04}).

\subsection{Optimization method}
The optimization (Opt.) method proposed by \citeauthor{whe00}
(\citeyear{whe00}) and developed by \citeauthor{wie04}
 (\citeyear{wie04}) consists in minimizing a joint measure for the
normalized Lorentz force and the divergence of the field, given by
the function,
\begin{equation}
\label{opit} L = \int_{V}\omega(x,y,z)[B^{-2}|(\nabla \times
\textbf{B}\times \textbf{B}) |^{2}+|\nabla\cdot
\textbf{B}|^{2}]d^{3}x,
\end{equation}
where $\omega(x,y,z)$ is a weighting function related position. It
is clear that (for $w > 0$) the force-free equations are fulfilled
when $L$ is equal to zero. This method involves minimizing $L$ by
optimizing the solution function $\textbf{B}(x, t)$ through states
that are increasingly force- and divergence-free, where $t$ is an
artificial time-like parameter introduced.

\subsection{Semi-analytical NLFF Field}
\citeauthor{low90} (\citeyear{low90}) have presented a class of
axis-symmetric NLFF fields in the spherical coordinate system. These
NLFF field can be used to test the performance of field
extrapolation by shifted under two steps of Cartesian coordinate
system transformation, rotating by an angle $\phi$ around the y-axis
and moving the origin to a distance $l$ along the z-axis. A special
class of NLFF fields denoted $\textbf{B}$($n$, $m$) can be obtained
in the spherical coordinate system. For different $n$ and $m$, it
can give different distributions of magnetic field, which meet the
requirements of divergence-free and force-free equations. Then we
should specify two parameters $l$ and $\phi$ for coordinate
transformation. In this paper we choose such two classic NLFF fields
as semi-analytical fields:

SAF1: the semi-analytical field with $n = 1, m = 1$, $l =0.3$, and
$\phi = \dfrac{\pi}{4}$, set $x \in [-0.5,0.5]$, $y \in [-0.5,0.5]$
and $z \in [0, 1]$ in the Cartesian coordinate system.

SAF2: the semi-analytical field with $n = 3, m = 1$, $l =0.3$, and
$\phi = \dfrac{4\pi}{5}$, set $x \in [-0.5,0.5]$, $y \in [-0.5,0.5]$
and $z \in [0, 1]$ in the Cartesian coordinate system.

The mesh is 64 $\times$ 64 $\times$ 64 for those two fields. Because
$\alpha$ will be studied mainly, the values of analytical $\alpha$
($\alpha_{Ana.}$) of SAF1 and SAF2 are calculated ({\it cf.}
\citeauthor{low90}, \citeyear{low90}) and saved in 3D position.

\section{Results}
\label{S-Result}
\subsection{SAF1}
\begin{figure}


   \centerline{\includegraphics[width=0.5\textwidth,clip=]{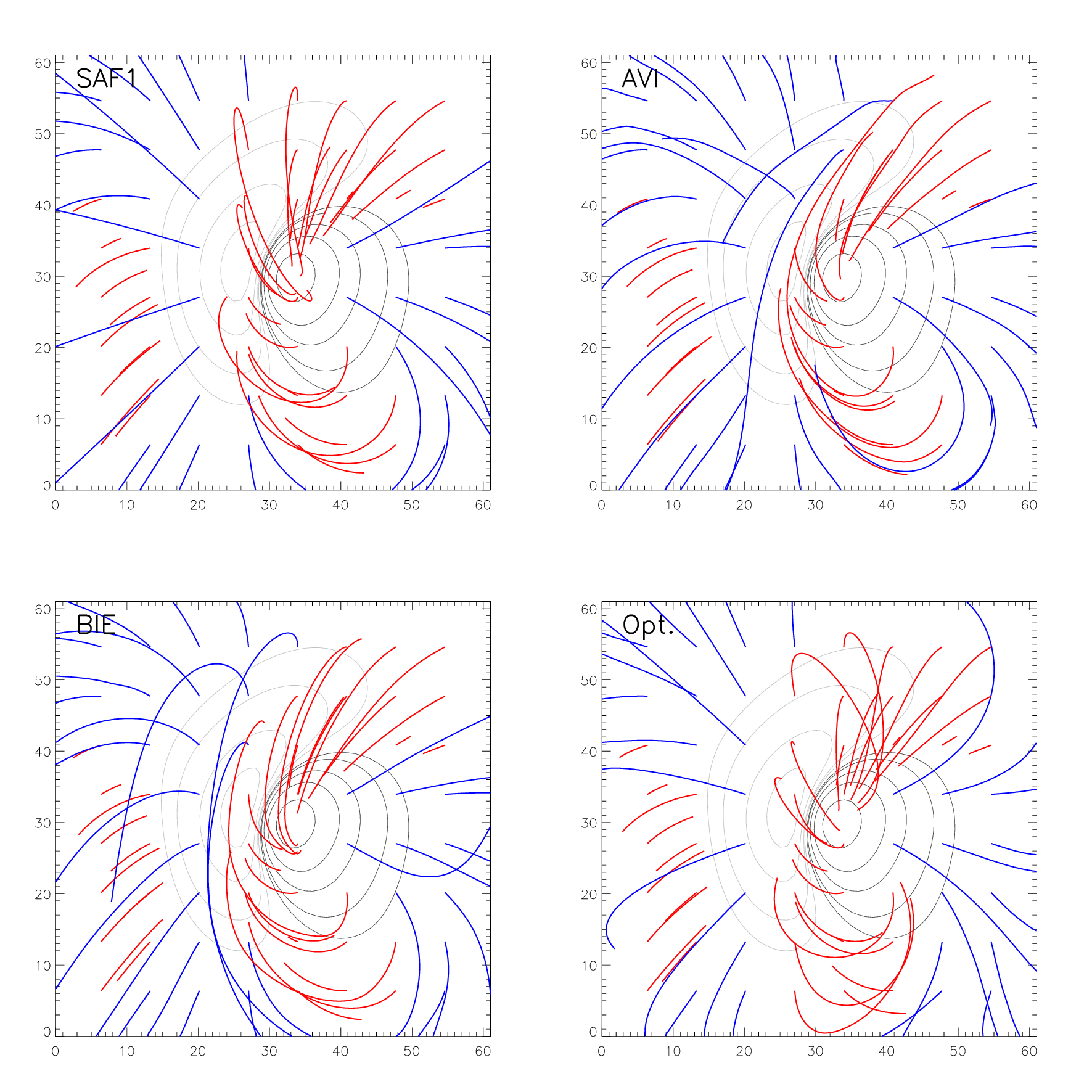}}

   \caption{The magnetic field lines for each NLFF fields (SAF1, AVI, BIE and Opt. methods are labeled).
} \label{Fig1}
\end{figure}

Figure~\ref{Fig1} shows the field lines of SAF1, AVI ,BIE and Opt.
fields. It can be found that the distributions of field lines of
extrapolated fields (AVI, BIE or Opt. field) are basically
consistent with that of SAF1, and the field lines of AVI and BIE
fields are very similar. Hence the structure of spatial field
obtained from field extrapolation can present for that of
semi-analytical field at some extent. In order to  evaluate the
performance of extrapolated field, we calculate the average
($\langle|F|\rangle$) and maximum ($F_{max}$) Lorentz-force
($\textbf{J}$~$\times$~$\textbf{B}/\mu_{0}$, $\mu_{0}$ is
permeability) and $\sigma_{j}$, defined by equation (\ref{sigma_j})
same as \citeauthor{sch06} (\citeyear{sch06}), to test the extent of
force-free for each extrapolated field. The results of these
physical quantity are given in table~1 for SAF1 and the
corresponding extrapolated fields, here the pixel size is assumed 1
arcsec. It can be found that angles between
$\textbf{J}$~and~$\textbf{B}$ of Opt. field (comparable to SAF1 at
some extent) are smaller than those of AVI and BIE fields.

\begin{equation}
\label{sigma_j}
 \sigma_{J} =\left(\sum_{i}\dfrac{|\textbf{J}_{i}\times \textbf{B}_{i}|}{B_{i}}\right)/ \sum_{i}J_{i},
\end{equation}

\begin{table}
\caption{The results of the average ($\langle|F|\rangle$) and
maximum ($F_{max}$) Lorentz-force and $\sigma_{j}$ for SAF1 and the
corresponding extrapolated fields. \label{tabsaf1ff}}
\begin{tabular}{lccccrr}
\hline
                  &$\langle|F|\rangle$ &       $F_{max}$  &                   $\sigma_{J}$ \\
            &($G^{2}/M$)$\times$ $10^{-14}$      &($G^{2}/M$)$\times$ $10^{-12}$    & \\
\hline
     SAF1         &2.2              &0.8              &0.14     \\
\hline
     AVI          &5.6              &9.6              &0.72         \\
\hline
     BIE          &3.7              &5.4              &0.45   \\
\hline
    Opt.          &2.3              &2.4              &0.21    \\
\hline
\end{tabular}
\end{table}

\begin{figure}
\centerline{\includegraphics[width=0.4\textwidth]{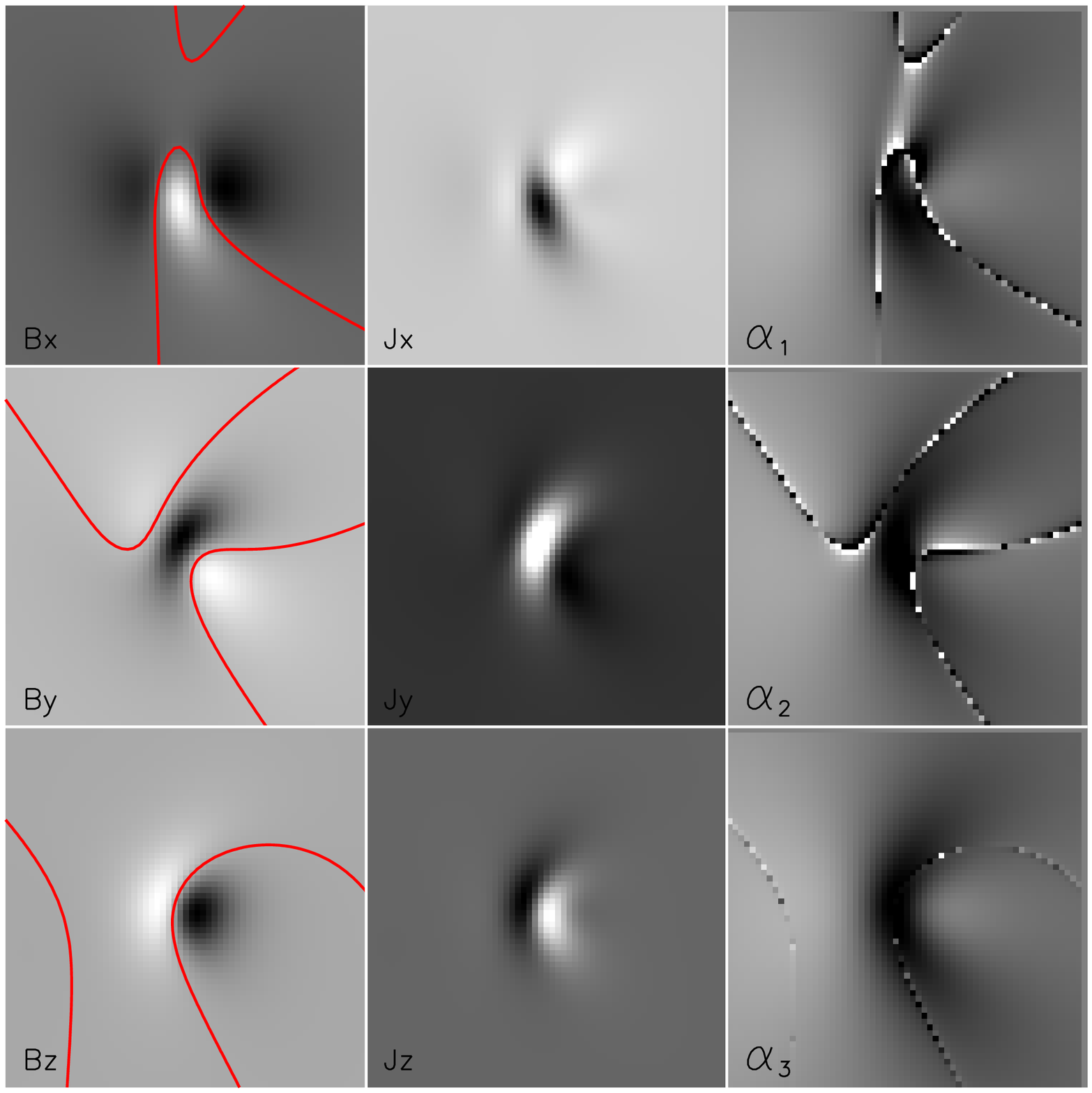}}
   \caption{The images of $\alpha_{1}$, $\alpha_{2}$ and $\alpha_{3}$,
   the components of magnetic field
    and current ($J_{x}$, $J_{y}$, $J_{z}$) labeled in each frame for SAF1 at $z=0$.}
\label{Fig2}
\end{figure}

Because a 3D field is available, force-free factor $\alpha$ can be
calculated individually from the formulas
(\ref{alpha1})-(\ref{alpha3}), which are obtained by expanding
Equation (\ref{wlfc}). Because $\alpha$ is the scalar depended on
spatial position, $\alpha_{1}$, $\alpha_{2}$ and $\alpha_{3}$ should
be the same for a given position. Figure~\ref{Fig2} shows the images
of $\alpha_{1}$, $\alpha_{2}$ and $\alpha_{3}$, magnetic field
components and current components ($J_{x}$, $J_{y}$, $J_{z}$)
labeled for SAF1 at $z=0$, where the red lines on each grey-scale
map of magnetic field component are the neutral line of each
component labeled. Because $\alpha$ is calculated from the
semi-analytical field, this figure demonstrates that there will be
unavoidable calculation errors when $\alpha$ is calculated from
magnetic field. The evident calculation errors are most likely
located near where the magnetic field components reversed, which are
consistent among $\alpha_{1}$, $\alpha_{2}$ and $\alpha_{3}$. From
the formulas (\ref{alpha1})-(\ref{alpha3}), it can be seen that
there are singularity problem when magnetic field component
approaches to zero, hence the evident calculation errors correspond
to the neutral line of magnetic field components (red lines
labeled). Since $\alpha$ is a scalar and the function of spatial
position, it is reasonable to combine $\alpha_{1}$, $\alpha_{2}$ and
$\alpha_{3}$ to minimize the calculation errors.

\begin{equation}
\label{alpha1} \alpha_{1}=J_{x}/B_{x},  \hspace{5ex}
J_{x}=\left(\dfrac{\partial B_{z}}{\partial y}-\dfrac{\partial
B_{y}}{\partial z}\right)\bigg /B_{x},
\end{equation}
\begin{equation}
\label{alpha2} \alpha_{2}=J_{y}/B_{y}, \hspace{5ex}
J_{y}=\left(\dfrac{\partial B_{x}}{\partial z}-\dfrac{\partial
B_{z}}{\partial x}\right)\bigg/B_{y},
\end{equation}
\begin{equation}
\label{alpha3} \alpha_{3}=J_{z}/B_{z}, \hspace{5ex}
J_{z}=\bigg(\dfrac{\partial B_{y}}{\partial x}-\dfrac{\partial
B_{x}}{\partial y}\bigg)\bigg/B_{z}.
\end{equation}

\begin{figure}

   \centerline{\includegraphics[width=0.5\textwidth,clip=]{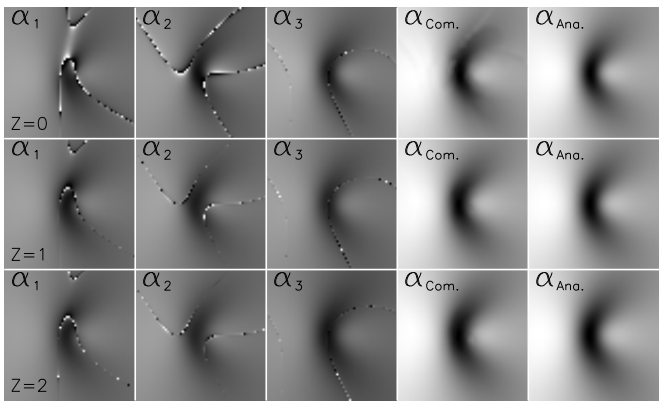}}
   \caption{The images of $\alpha_{1}$, $\alpha_{2}$, $\alpha_{3}$,
   $\alpha_{Com.}$ and $\alpha_{Ana.}$ at $z=0$, $z=1$ and $z=2$ for SAF1.
.} \label{Fig3}
\end{figure}

In order to compare $\alpha$, the combined value from $\alpha_{1}$,
$\alpha_{2}$ and $\alpha_{3}$ is also studied in this work. The
method to realize this is that: to get the average of two values of
$\alpha$ ($\alpha_{1}$, $\alpha_{2}$ or $\alpha_{3}$) whose values
are closed, which means that for a given position $\alpha$ deduced
from one formula (\ref{alpha1}, \ref{alpha2} or \ref{alpha3}) is
give up forcibly. Figure~\ref{Fig3} shows the process to combine
$\alpha$ reasonably. It shows $\alpha_{1}$, $\alpha_{2}$ and
$\alpha_{3}$, the combined $\alpha$ ($\alpha_{Com.}$) and analytical
$\alpha$ ($\alpha_{Ana.}$) of SAF1 at $z=0$, $z=1$ and $z=2$. It is
found that there are evident calculation errors for $\alpha_{1}$,
$\alpha_{2}$ and $\alpha_{3}$. The correlation coefficients between
$\alpha_{1,2,3,Com.}$ and $\alpha_{Ana.}$ are given in table~2.
However $\alpha_{Com.}$ is improved evidently, $\alpha_{Com.}$ match
very well with $\alpha_{Ana.}$, the correlation coefficients between
$\alpha_{Com.}$and $\alpha_{Ana.}$ are 0.99. Therefore the process
to combine $\alpha$ is valid for SAF1 (note this validation is not
suitable for all cases of extrapolated field, but it can be valid on
the whole, which will present next section).

\begin{table}

\caption{The correlation coefficients between $\alpha_{1,2,3,Com.}$
and $\alpha_{Ana.}$ for SAF1. \label{tab1}}
\begin{tabular}{clrrrrrrrrrr}
\hline
                  &$\alpha_{Com.}$ &  $\alpha_{1}$  &   $\alpha_{2}$ & $\alpha_{3}$ \\
\hline
     z=0          &0.99              &0.97              &0.93         &0.90  \\
\hline
     z=1          &0.99              &0.92              &0.89         &0.91 \\
\hline
     z=2          &0.99               &0.97              &0.90        &0.93  \\
\hline
\end{tabular}
\end{table}

\begin{figure}

   \centerline{\includegraphics[width=.5\textwidth,clip=]{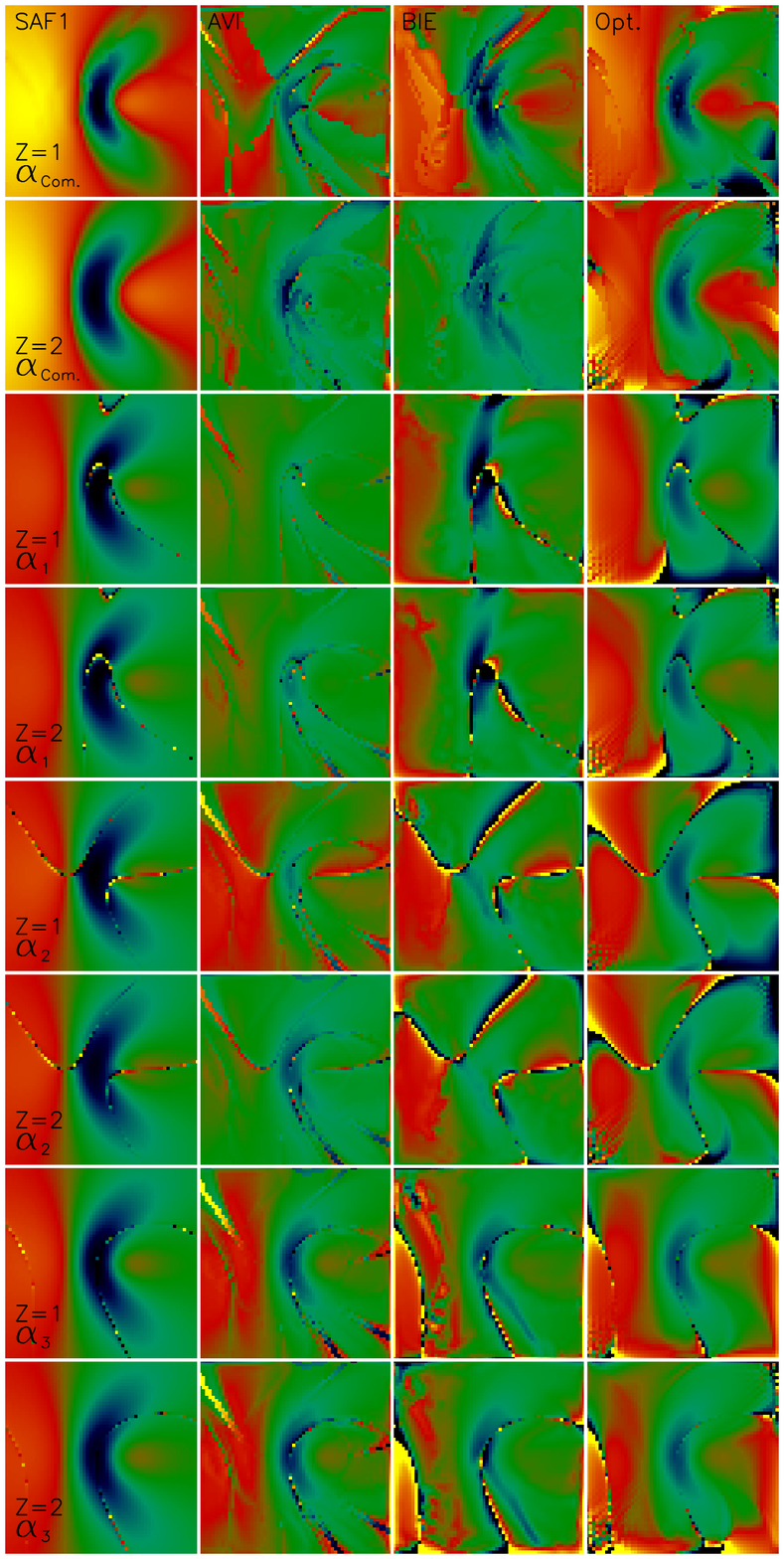}}
   \caption{The images of $\alpha_{Com.,1,2,3}$ deduced from extrapolated
   fields at $z=1$ and $z=2$. (SAF1)}\label{Fig4}
\end{figure}

Figure~\ref{Fig4} shows the images of $\alpha_{1,2,3,Com.}$ deduced
from SAF1, AVI, BIE and Opt. fields at $z=1$ and $z=2$. There are
evident deviations of $\alpha$ between extrapolated field and
semi-analytical field. The correlation coefficients of
$\alpha_{1,2,3,Com.}$ between SAF1 and each extrapolated fields are
given in table~3. This low relation of $\alpha$s between
extrapolated field and SAF1 is caused mainly by the deviations
between extrapolated field and SAF1, since from the above analysis
it is found that the correlation coefficients between
$\alpha_{Com.}$ and $\alpha_{Ana.}$ are already greater than 99\%
for SAF1.
\begin{table}
\caption{The correlation coefficients of $\alpha_{1,2,3,Com.}$
between SAF1 and the corresponding extrapolated fields.
\label{tab2}}
\begin{tabular}{lccccccccccccccccccc}
\hline~
                           &AVI    &BIE    &Opt.  \\       
\hline\hline
     $\alpha_{Com.}(z=1)$ &0.36   &0.49  &0.67     \\      
\hline
     $\alpha_{Com.}(z=2)$ &0.42   &0.41  &0.59     \\      
\hline
\hline
     $\alpha_{2}(z=1)$ &0.43   &0.59  &0.62         \\     
\hline
     $\alpha_{2}(z=2)$ &0.46   &0.46  &0.54         \\     

\hline\hline
     $\alpha_{1}(z=1)$   &0.33   &0.39   &0.55\\

\hline
     $\alpha_{1}(z=2)$   &0.38    &0.41  &0.56\\

\hline\hline
     $\alpha_{3}(z=1)$   &0.65   &0.45   &0.53\\
\hline
      $\alpha_{3}(z=2)$   &0.58    &0.52  &0.63\\
\hline
\end{tabular}
\end{table}

\begin{equation}\label{rsd}
RSD=\dfrac{1}{\langle\alpha\rangle}\sqrt{\dfrac{1}{N-1}\sum_{i=0}^{N-1}(\alpha_{i}-\langle\alpha\rangle)^{2}}.
\end{equation}

\begin{figure}
   \centerline{\includegraphics[width=0.5\textwidth,clip=]{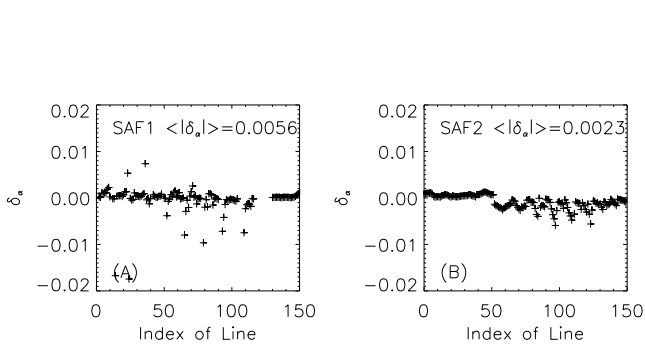}}
   \caption{RSD ($\delta_{\alpha}$) of $\alpha_{Ana.}$ calculated from
   selected field lines for two semi-analytical fields (labeled).}
\label{Fig5}
\end{figure}

As for the NLFF field, force-free factor $\alpha$ should be a
constant along one special field line. Hence a physical quantity:
relative standard deviation (RSD),  which can estimate the deviation
of $\alpha$ along field lines is studied. RSD is defined by formula
($\ref{rsd}$), where N is the number of points calculated along
field line. Figure~\ref{Fig5}(A) shows RSD of $\alpha_{Ana.}$ along
each field lines for SAF1. Because the values of RSD approach zero
($\langle |RSD|\rangle$ is only 0.0056 and the points is very
concentrated, which can give the conclusion that the field lines can
be traced and the deviation of $\alpha_{Ana.}$ of SAF1 is
negligible. It means that SAF1 satisfies the force-free equations
very well.

\begin{figure}
   \centerline{\includegraphics[width=0.5\textwidth,clip=]{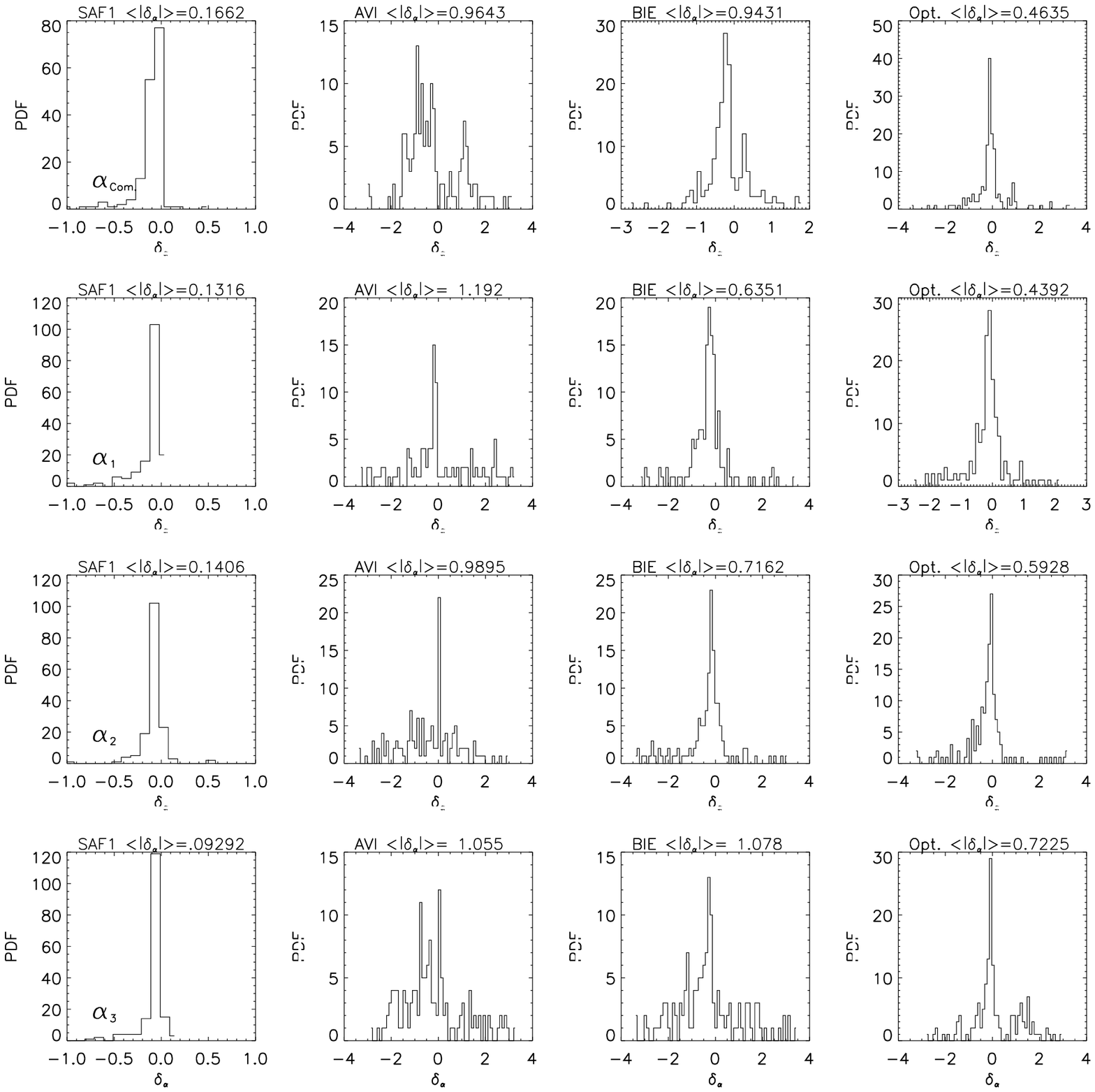}}
   \caption{The PDF of RSD ($\delta_{\alpha_{Com.,1,2,3}}$) calculated from some
   selected field lines for each NLFF field. (SAF1)}
\label{Fig6}
\end{figure}

\begin{figure}
   \centerline{\includegraphics[width=0.5\textwidth,clip=]{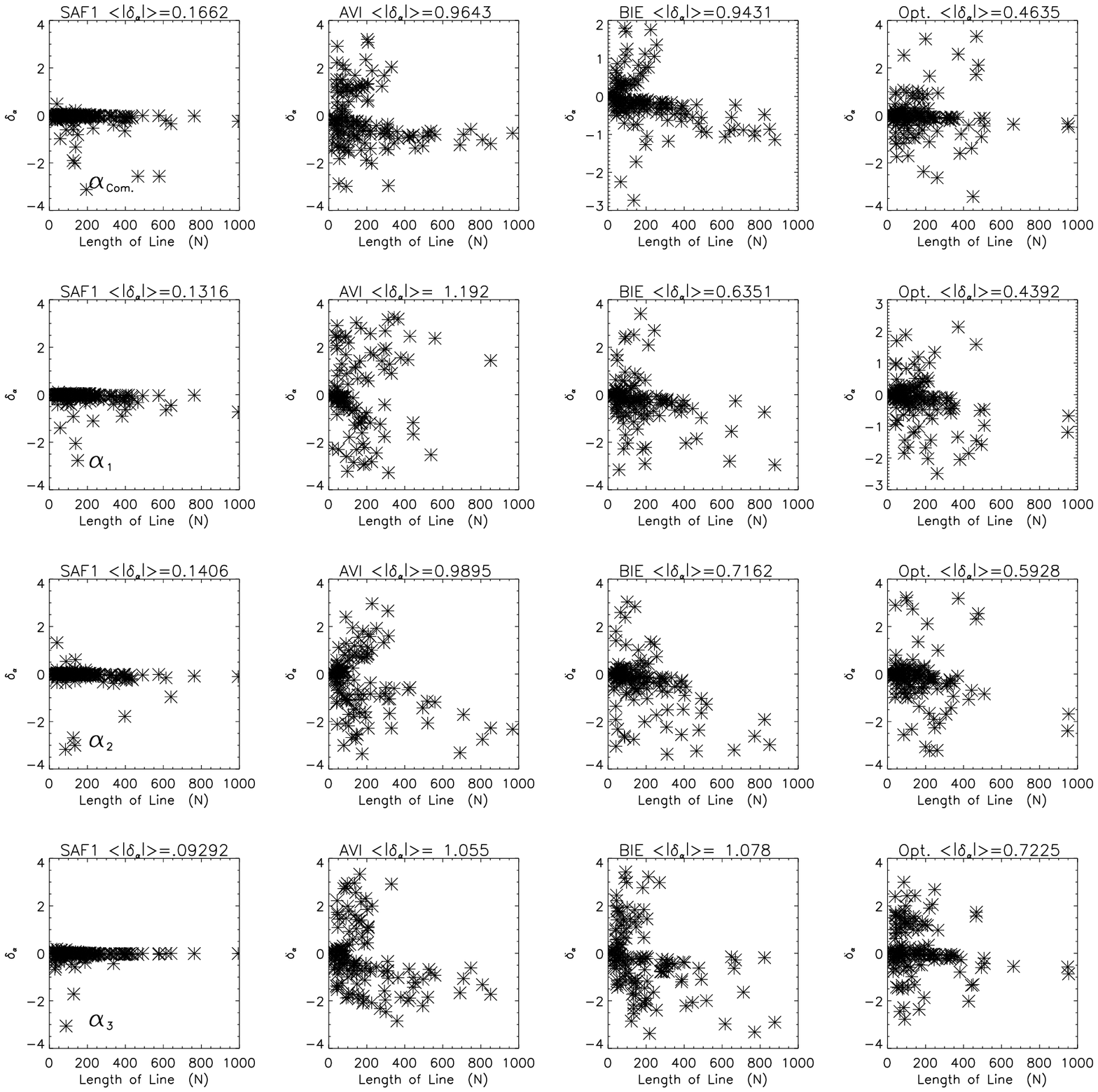}}
   \caption{The scatter diagram of RSD ($\delta_{\alpha_{Com.,1,2,3}}$) vs the length of field line
   calculated from some selected field lines for each NLFF field. (SAF1)}

\label{Fig7}
\end{figure}

In Figure~\ref{Fig6}, the possibility function (PDF) of RSD of
$\alpha_{1,2,3,Com.}$ along some selected field lines are plotted
for SAF1, AVI, BIE and Opt. fields. The reason for RSD of SAF1 are
plotted is that they can give an estimation on calculation errors.
It can be found that PDF of RSD of $\alpha$ is very narrow and
$\langle |RSD|\rangle$ is 0.09-0.16 for SAF1, this value can be
taken as the calculation error of $\langle |RSD|\rangle$. For
extrapolated NLFF fields, there are evident deviations of $\alpha$
along field lines since PDF of RSD of $\alpha$ is not concentrated
and $\langle |RSD|\rangle$ of $\alpha$ along some selected field
lines is 0.96-1.19, 0.63-1.07 and 0.43-0.72 for AVI, BIE and Opt.
field, respectively. Additionally, the validation of combined is not
evident for some cases,for example BIE extrapolated field (its RSD
of
 $\alpha_{Com.}$ is not the smallest one). In Figure~\ref{Fig7},
 RSD vs field line length are plotted for
SAF1, AVI, BIE and Opt. fields in order to find whether or not RSD
are depend on field line length, where field line length was
indicated by the number of points calculated along each field line.
It can be found that RSD dose basically not depend on field line
length.
\subsection{SAF2}
\begin{figure}
   \centerline{\includegraphics[width=0.5\textwidth,clip=]{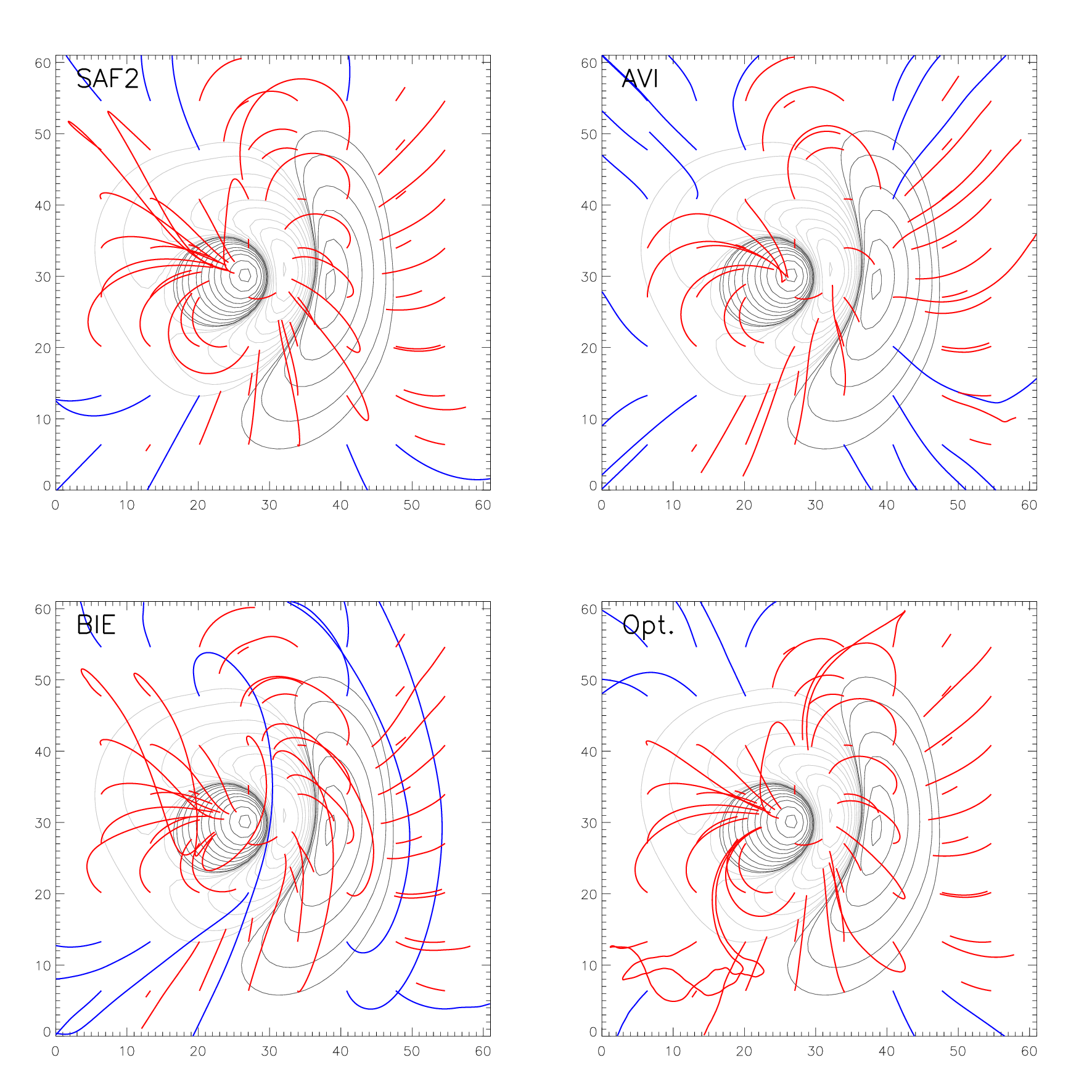}}

   \caption{The magnetic field lines for each NLFF fields
   (SAF2, AVI, BIE and Opt. methods are labeled)
} \label{Fig8}
\end{figure}

Figure~\ref{Fig8} shows the field lines of SAF2, AVI, BIE and Opt.
fields. Same as Figure~\ref{Fig1} it can be found that although
there are some fine differences of field lines among these fields,
the distributions of field lines of extrapolated fields basically
match that of SAF2 at large scale. Like SAF1, we also calculate
$\langle F\rangle$, ~$F_{max}$ and $\sigma_{j}$ for each
extrapolated field corresponding to SAF2. The results are given in
table~4, here the pixel size is also assumed 1 arcsec. It can be
found that for SAF2 angles between $\textbf{J}$~and~$\textbf{B}$ are
large than those for SAF1, even for semi-analytical field. The
$\langle F\rangle$~and~$F_{max}$ differences among extrapolated
fields are negligible, however the $\sigma_{j}$ of Opt. field is
better than those of AVI and BIE fields.

\begin{table}\label{tabsaf2ff}
\caption{The results of the average ($\langle F\rangle$) and maximum
($F_{max}$) Lorentz-force and $\sigma_{j}$ for SAF2 and the
corresponding extrapolated fields. }
\begin{tabular}{cccccrr}
\hline
                  &$\langle F\rangle$ &       $F_{max}$  &                   $\sigma_{J}$ \\
            &($G^{2}/M$)$\times$ $10^{-12}$      &($G^{2}/M$)$\times$ $10^{-11}$    &  \\
\hline
     SAF2         &1.4              &13.6              &0.30      \\
\hline
     AVI          &9.3              &25.7               &0.69         \\
\hline
     BIE          &7.7              &32.4               &0.71   \\
\hline
    Opt.          &1.8              &2.6              &0.31    \\
\hline
\end{tabular}
\end{table}

\begin{figure}
\centerline{\includegraphics[width=0.4\textwidth]{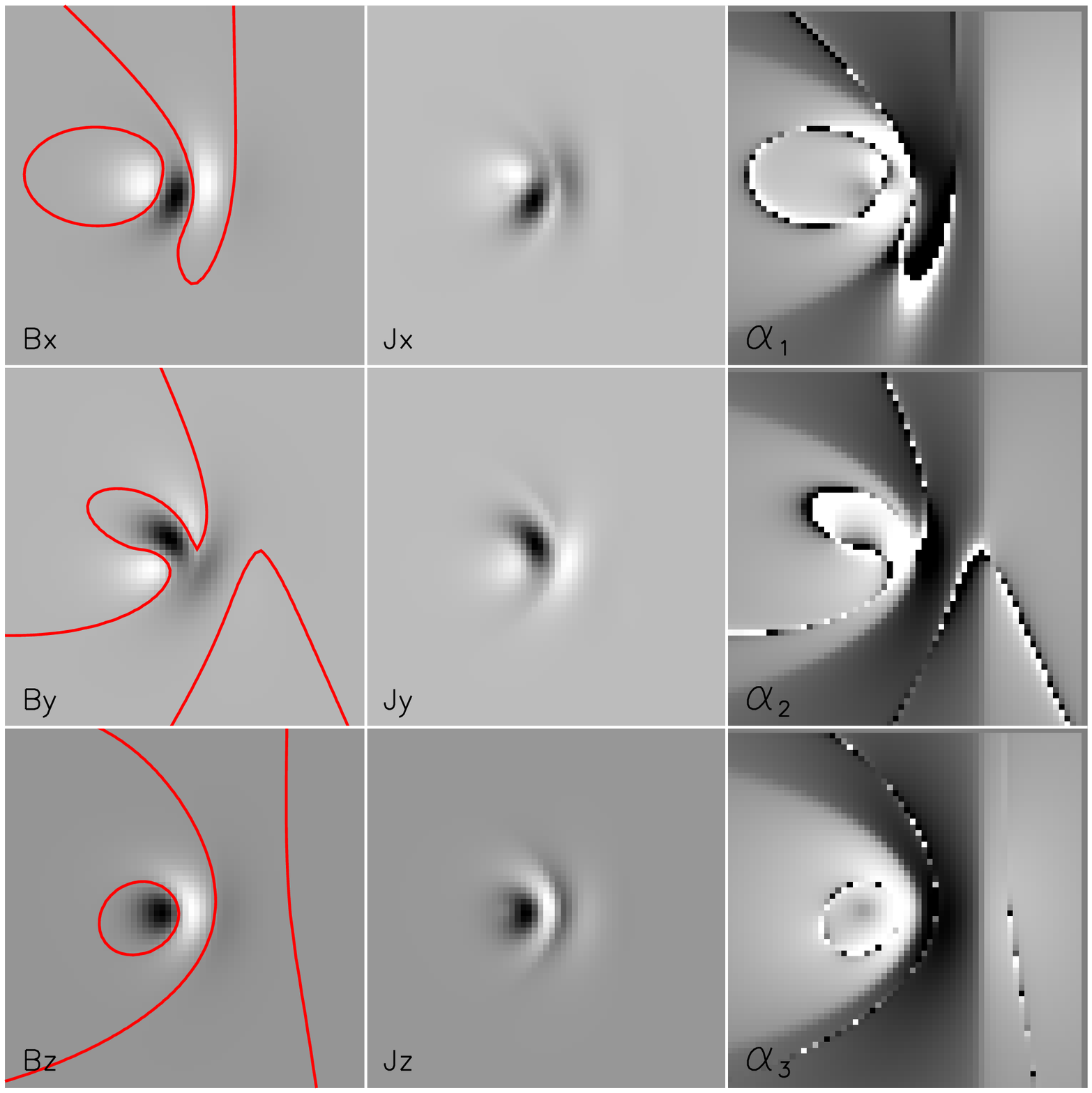}}
   \caption{The images of $\alpha_{1}$, $\alpha_{2}$ and $\alpha_{3}$,
   the components of magnetic field
    and current ($J_{x}$, $J_{y}$, $J_{z}$) labeled in each frame for SAF2 at $z=0$.}

\label{Fig9}
\end{figure}

\begin{figure}
   \centerline{\includegraphics[width=0.5\textwidth,clip=]{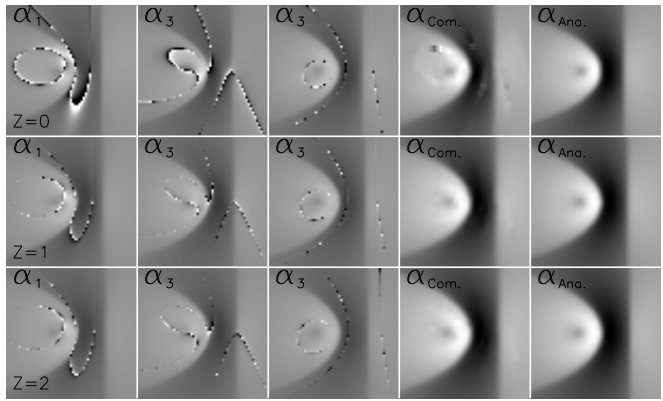}}
   \caption{The images of $\alpha_{1}$, $\alpha_{1}$ and $\alpha_{1}$,
    $\alpha_{Com.}$ and $\alpha_{Ana.}$ at $z=0$, $z=1$ and $z=2$ for SAF2.}
\label{Fig10}
\end{figure}

Same as Figure~\ref{Fig2} for SAF1, Figure~\ref{Fig9} shows the
images of $\alpha_{1}$, $\alpha_{2}$ and $\alpha_{3}$, components of
magnetic field and current for SAF2 at $z=0$. This also demonstrates
that calculation errors are mainly located near where the magnetic
field components reversed. Like Figure~\ref{Fig3} for SAF1,
Figure~\ref{Fig10} shows $\alpha_{1}$, $\alpha_{2}$, $\alpha_{3}$,
$\alpha_{Com.}$ and $\alpha_{Ana.}$ of SAF2 at $z=0$, $z=1$ and
$z=2$. Through calculation we get the correlation coefficients
between $\alpha_{1,2,3,Com.}$ and $\alpha_{Ana.}$, which are shown
in table~5. It shows that $\alpha_{Com.}$ evidently improve the
computational precision of $\alpha$ for SAF2.
\begin{table} \label{tab3}
\caption{ he correlation coefficients between $\alpha_{1,2,3,Com.}$
and $\alpha_{Ana.}$ for SAF2.}
\begin{tabular}{clrrrrrrrrrr}
\hline
                  &$\alpha_{Com.}$ &  $\alpha_{1}$  &   $\alpha_{2}$ & $\alpha_{3}$ \\
\hline
     z=0          &0.96              &0.96              &0.92         &0.92  \\
\hline
     z=1          &0.99              &0.94              &0.94         &0.94 \\
\hline
     z=2          &0.99               &0.93              &0.97        &0.97  \\
\hline
\end{tabular}
\end{table}

\begin{figure}
   \centerline{\includegraphics[width=.5\textwidth,clip=]{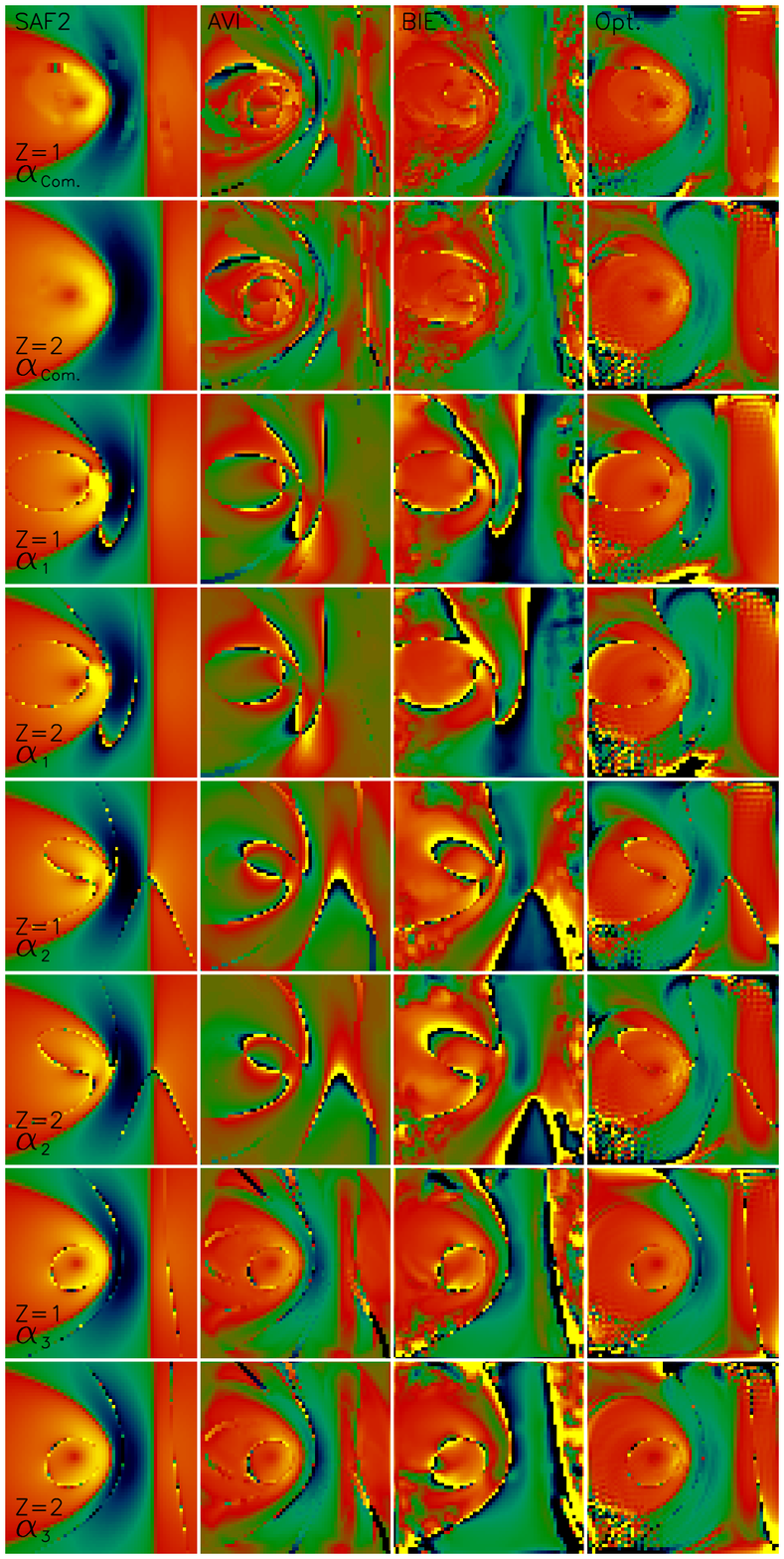}}
   \caption{The images of $\alpha_{Com.,1,2,3}$ deduced from extrapolated fields
    at $z=1$ and $z=2$. (SAF2)}
\label{Fig11}
\end{figure}

Figure~\ref{Fig11} shows the images of $\alpha_{1,2,3,Com.}$ deduced
from SAF2, BIE, AVI and Opt. fields at $z=1$ and $z=2$. Comparing
the results of SAF1, the consistency between $\alpha$s of
extrapolated field and SAF2 is improved at some extent. The
correlation coefficients of $\alpha_{1,2,3,Com.}$ between SAF1 and
the corresponding extrapolated fields are given in table~6. This
relative low relation of $\alpha$s between extrapolated field and
SAF2 is also caused by the deviations extrapolated field and
semi-analytical field.

\begin{table}\label{tab4}
\caption{The correlation coefficients of $\alpha_{1,2,3,Com.}$
between SAF1 and the corresponding extrapolated fields. }
\begin{tabular}{lccccccccccccccccccc}
\hline~
                           &AVI    &BIE    &Opt. \\     
\hline\hline
     $\alpha_{Com.}(z=1)$ &0.54   &0.77  &0.72   \\     
\hline
     $\alpha_{Com.}(z=2)$ &0.59   &0.73  &0.68   \\     
\hline\hline
\hline
     $\alpha_{2}(z=1)$ &0.48   &0.71  &0.67     \\        
\hline
     $\alpha_{2}(z=2)$ &0.54   &0.64  &0.54     \\     

\hline\hline
     $\alpha_{1}(z=1)$   &0.56   &0.65   &0.71  \\
\hline
     $\alpha_{1}(z=2)$   &0.51    &0.63  &0.66   \\

\hline\hline
     $\alpha_{3}(z=1)$   &0.65   &0.58   &0.68  \\
\hline
     $\alpha_{3}(z=2)$   &0.58    &0.55  &0.65   \\
\hline
\end{tabular}
\end{table}

Figure~\ref{Fig5}(B) shows RSD of $\alpha_{Ana.}$ for SAF2, where
$\langle |RSD|\rangle$ is 0.0023 and the points is also very
concentrated, which also indicate that deviation of $\alpha_{Ana.}$
of SAF2 is negligible. It means that SAF2 also satisfies force-free
equations very well.

\begin{figure}
   \centerline{\includegraphics[width=0.5\textwidth,clip=]{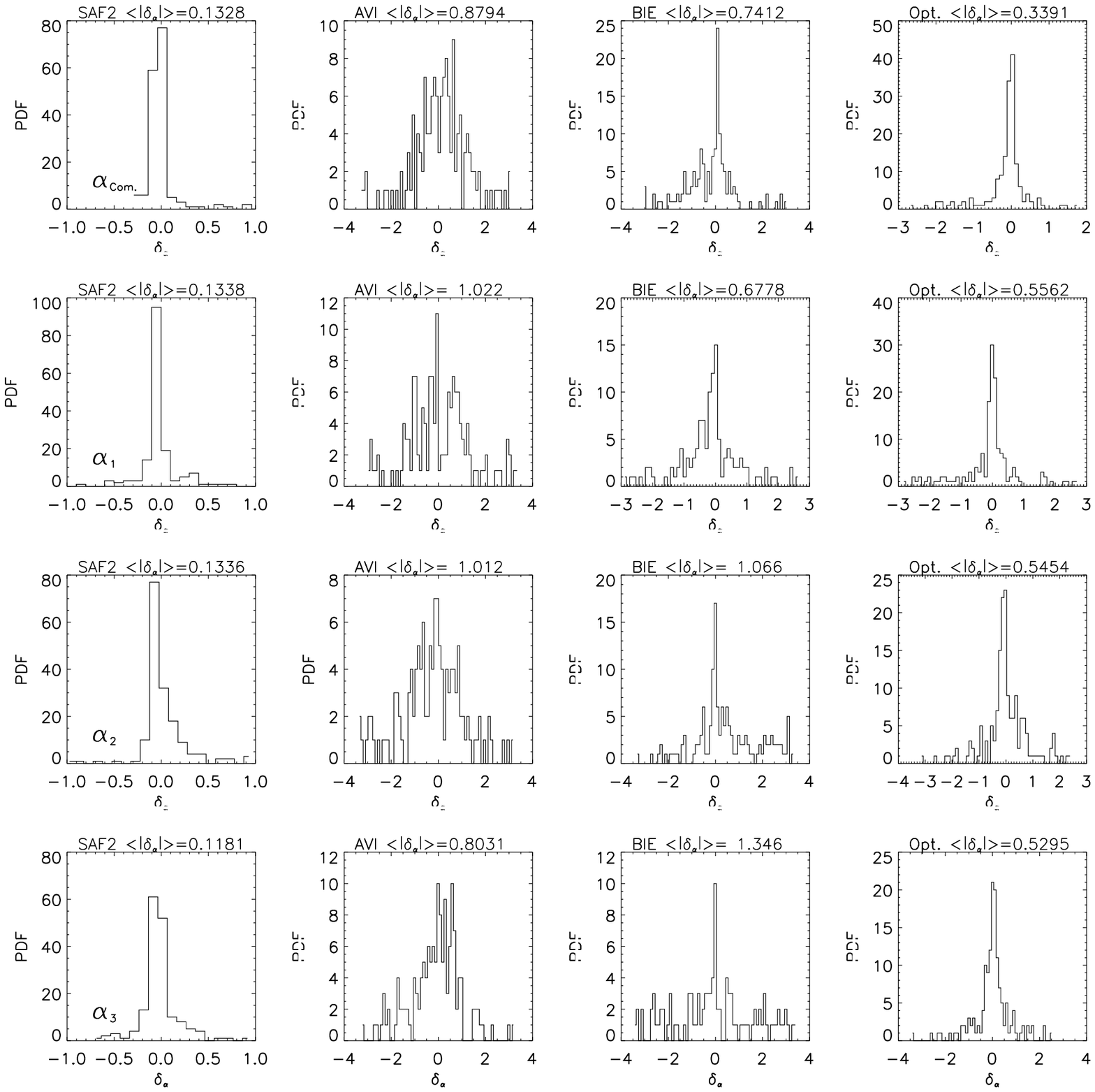}}
   \caption{The PDF of RSD ($\delta_{\alpha_{Com.,1,2,3}}$) calculated from some
   selected field lines for each NLFF field. (SAF2)}
\label{Fig12}
\end{figure}

\begin{figure}
   \centerline{\includegraphics[width=0.5\textwidth,clip=]{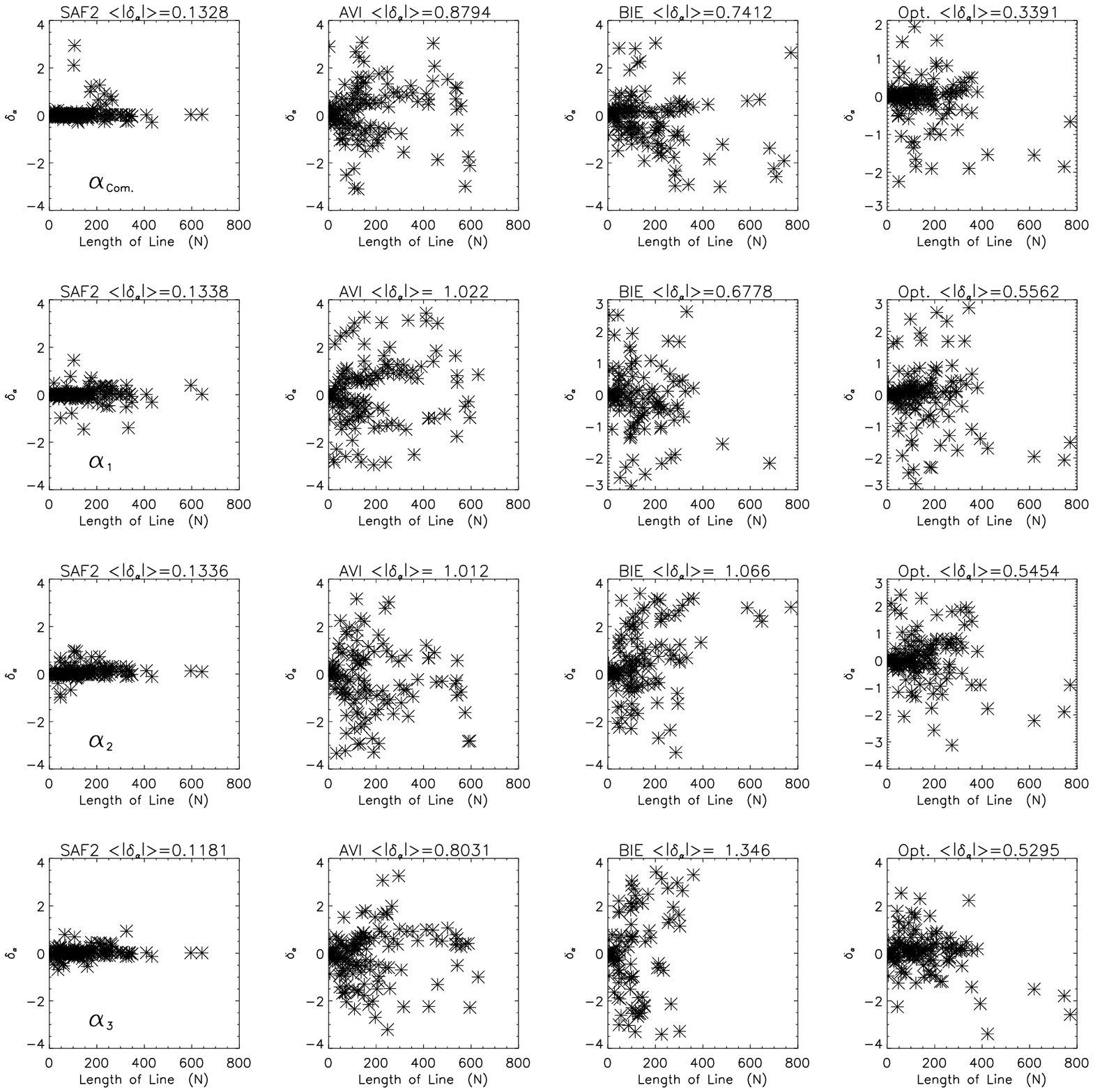}}
   \caption{The scatter diagram of RSD ($\delta_{\alpha_{Com.,1,2,3}}$) vs the length of field
   line calculated from some selected field lines for each NLFF field. (SAF2)}
\label{Fig13}
\end{figure}

Same as Figure~\ref{Fig6}, PDF of RSD of $\alpha$ along some
selected field lines are plotted in Figure~\ref{Fig12} for SAF2,
BIE, AVI and Opt. fields. From this figure, it can be found that
calculation errors denoted by RSD, that deduced from SAF2, is
smaller than those of SAF1. $\langle |RSD|\rangle$ is 0.13-0.11,
0.80-1.02, 0.67-1.34 and 0.33-0.55 for SAF2, AVI, BIE and Opt.
fields, respectively, which are better than those of SAF1. However
the PDFs of RSD are also very wide for these extrapolated NLFF
fields. RSD vs field line length for SAF2, AVI, BIE and Opt. fields
are plotted in Figure~\ref{Fig13}, it can give the same results
consistent with those of SAF1, that RSD do not depend on field line
length either.

\section{Discussions and Conclusions}
\label{S-Conl}

 In this paper, force-free factor $\alpha$ of NLFF
field was mainly studied. The aim is to find, to what extent,
$\alpha$ along a given field line can keep a constant, and to give
an error estimation on $\alpha$ calculated from magnetic field.

Through analysis, it is found that there are unavoidable calculation
errors for deducing $\alpha$ from magnetic field, the calculation
errors are most likely to locate near where magnetic field
components are reversed. $\langle |RSD|\rangle$ of $\alpha$ along
selected field lines is about 0.1 $\sim$ 0.2 for semi-analytical
fields, which can be considered as calculation errors of $\langle
|RSD|\rangle$ caused by computation completely.

It is found that there are obvious deviations on $\alpha$ of
extrapolated fields from that of semi-analytical fields. The results
of deviation of $\alpha$ along selected field lines are as follows:
For SAF1, $\langle |RSD|\rangle$ of $\alpha$ along selected field
lines are about 0.96-1.19, 0.63-1.07 and 0.43-0.72, for AVI, BIE and
Opt. fields, respectively. While for SAF2, they are about 0.13-0.11,
0.80-1.02, 0.67-1.34 and 0.33-0.55 for AVI, BIE and Opt. fields,
respectively. In both cases, it can be found that RSD of $\alpha$
along the selected field lines do not depend on field line length.

Since RSD is a criterion for the deviation from a perfect force-free
state (the lower values indicate a more accurate force-free state)
and RSD of Opt. field is less than those of AVI and BIE fields
basically, so the performance of Opt. extrapolated fields is
superior to other two extrapolated fields at some extent. In
addition, an interesting thing is that $\alpha_{Com.}$ is valid for
all case of Opt. field, which is the same as semi-analytical fields.
Previous studies mostly paid close attentions to the global or point
to point properties for extrapolated fiedl, even if these properties
are reasonable and acceptable, another property of force-free field
($\alpha$~ should be an constant along field line) may not be
satisfied well. For example, BIE method have given an constrain on
the angles between $\textbf{B}$~(calculated from Helmholtz equation)
and~$\textbf{J}$~(deduced from $\textbf{B}$) at each point, so the
reasonability of global performance may be improved on the whole,
but higher requirements ($\alpha$~ should be an constant along field
line) should be add for extrapolation method. For the extrapolation
errors of AVI method, there are two main effects, first is that it
reconstructs the field by two field terms; second is that the
singularity problems can not removal completely. In fact, the
initial condition of potential field for Opt. field may give better
results of RSD at some extent, however the superiority of Opt. field
can also be studied from other aspects, such as the correlation
coefficients of $\alpha$ between extrapolated field and
semi-analytical field and other global properties.

\acknowledgments This work was partly supported by the
 National Natural Science Foundation of China (Grant Nos. 10611120338,
 10673016, 10733020, 10778723, 11003025 and 10878016),
 National Basic  Research Program of China (Grant No. 2011CB8114001) and Important
 Directional Projects of Chinese Academy of Sciences (Grant No. KLCX2-YW-T04).





%

%


%

\begin{thebibliography}{}
\bibitem[\protect\citeauthoryear{{Aly}}{1989}]{aly89}
Aly, J.J.: 1989, \solphys{} \textbf{120}, 19.

\bibitem[\protect\citeauthoryear{{Amari \etal}}{1997}]{ama97}
Amari, T., Aly, J.J., Luciani, J.F., Boulmezaoud, T.Z., Mikic, Z.:
1997, \solphys{} \textbf{174}, 129.

\bibitem[\protect\citeauthoryear{{Amari \etal}}{2006}]{ama06}
Amari, T., Boulmezaoud, T.Z., Aly, J.J.: 2006, \aap{} \textbf{446},
691.

\bibitem[\protect\citeauthoryear{{Chodura and Schlueter}}{1981}]{cho81}
Chodura, R., Schlueter, A.: 1981, \jcp{} \textbf{41}, 68.


\bibitem[\protect\citeauthoryear{{Cuperman \etal}}{1990}]{cuper90}
Cuperman, S., Ofman, L., Semel, M.: 1990, \aap{} \textbf{230}, 193.

\bibitem[\protect\citeauthoryear{{D{\'e}moulin \etal}}{1992}]{demo92}
D{\'e}moulin, P., Cuperman, S., Semel, M.: 1992,  \aap{}
\textbf{236}, 351.

\bibitem[\protect\citeauthoryear{{DeRosa \etal}}{2009}]{sch09}
DeRosa, M.L., Schrijver, C.J., Barnes, G., Leka, K.D., Lites, B.W.,
Aschwanden, M.J., {\it et al.}: 2009, \apj{}  \textbf{696}, 1780

\bibitem[\protect\citeauthoryear{{He and Wang}}{2008}]{he08}
He, H., Wang, H.: 2008, \jgr{} \textbf{113}, A05S90

\bibitem[\protect\citeauthoryear{{He \etal}}{2011}]{he11}
He, H., Wang, H.,Yan, Y.: 2011, \jgr{} \textbf{116}, A01101

\bibitem[\protect\citeauthoryear{{Liu \etal}}{2011}]{lius11}
Liu, S., Zhang, H.Q., Su, J.T. and Song, M.T.: 2011, \solphys{}
\textbf{269}, 41


\bibitem[\protect\citeauthoryear{{Li, Yan, and Song}}{2004}]{li04}
Li, Z, Yan, Y.H., Song, G.: 2004, \mnras{} \textbf{347}, 1255


\bibitem[\protect\citeauthoryear{{Low and Lou}}{1990}]{low90}
Low, B.C., Lou, Y.Q.: 1990, \apj{} \textbf{352}, 343


\bibitem[\protect\citeauthoryear{{Mikic and McClymont}}{1994}]{mic94}
Mikic, Z.; McClymont, A. N.: 1994, in Solar Active Region Evolution:
Comparing Models with Observations, Vol68. ASP Conf. Ser., p.225.

\bibitem[\protect\citeauthoryear{{R{\'e}gnier \etal}}{2004}]{regnier04}
R{\'e}gnier, S., Amari, T.: 2004,  \aap{} \textbf{425}, 345.

\bibitem[\protect\citeauthoryear{{R{\'e}gnier \etal}}{2007}]{regnier07}
R{\'e}gnier, S., Priest, E. R.: 2007,  \aap{} \textbf{468}, 701.

\bibitem[\protect\citeauthoryear{{Roumeliotis}}{1996}]{rou96}
Roumeliotis, G.: 1996, \apj{} \textbf{473}, 1095

\bibitem[\protect\citeauthoryear{{Sakurai}}{1981}]{sak81}
Sakurai, T.: 1981, \solphys{} \textbf{69}, 343.

\bibitem[\protect\citeauthoryear{{Schrijver \etal}}{2006}]{sch06}
Schrijver, C.J., De Rosa, M. L., Metcalf, T. R., Liu, Y., McTiernan,
J., R{\'e}gnier, S., Valori, G., Wheatland, M. S., Wiegelmann, T.:
2006, \solphys{} \textbf{235}, 161.

\bibitem[\protect\citeauthoryear{{Song \etal}}{2006}]{son06}
Song, M.T., Fang, C., Tang, Y.H., Wu, S.T., Zhang, Y.A.: 2006,
\apj{} \textbf{649}, 1084.

\bibitem[\protect\citeauthoryear{{Valori \etal}}{2007}]{valo07}
Valori, G., Kliem, B., Fuhrmann, M.: 2007, \solphys{} \textbf{245},
263.

\bibitem[\protect\citeauthoryear{{Wheatland \etal}}{2000}]{whe00}
Wheatland, M.S., Sturrock, P.A., Roumeliotis, G.: 2000, \apj{}
\textbf{540}, 1150.

\bibitem[\protect\citeauthoryear{{Wiegelmann}}{2004}]{wie04}
Wiegelmann, T.: 2004, \solphys{} \textbf{219}, 87.


\bibitem[\protect\citeauthoryear{{Wiegelmann \etal}}{2006}]{wie06}
Wiegelmann, T., Inhester, B., Sakurai, T.: 2006, \solphys{}
\textbf{233}, 215

\bibitem[\protect\citeauthoryear{{Wu \etal}}{1990}]{wu90}
Wu, S.T., Sun, M.T., Chang, H.M., Hagyard, M.J., Gary, G.A.: 1990,
\apj{} \textbf{362}, 698.

\bibitem[\protect\citeauthoryear{{Yan and Li}}{2006}]{yan06}
Yan, Y., Li, Z.: 2006, \apj{} \textbf{638}, 1162.

\bibitem[\protect\citeauthoryear{{Yan and Sakurai}}{2000}]{yan00}
Yan, Y., Sakurai, T.: 2000, \solphys{} \textbf{195}, 89.
\end{thebibliography}

%

\end{document}